\documentclass[prb, twocolumn, showpacs, amsmath, amssymb]{revtex4}

\usepackage{graphicx}
\usepackage{dcolumn}
\usepackage{bm}

\begin{document}

\preprint{APS/123-QED}

\title{Nonmagnetic impurity effects in MgB$_{2}$}

\author{Koichi Watanabe}
\affiliation{Division of Physics,  Hokkaido University,  Sapporo 060-0810,  Japan}
\author{Takafumi Kita}
\affiliation{Division of Physics,  Hokkaido University,  Sapporo 060-0810,  Japan}

\date{\today}

\begin{abstract}
We study nonmagnetic impurity effects in MgB$_{2}$ using the quasiclassical equations of
superconductivity for a weak-coupling two-band model. 
Parameters in the model are fixed so as to reproduce experiments on MgB$_{2}$
as closely as possible.
The quasiparticle density of states and the specific heat are calculated
for various values of the interband impurity scattering.
The density of states changes gradually from a two-gap structure into the conventional
single-gap structure as the interband scattering increases.
It is found that the excitation threshold is not a monotonic function of
the interband scattering. 
Calculated results for the specific heat are in good agreements with experiments
on samples after irradiation. 
\end{abstract}

\pacs{Valid PACS appear here}
\maketitle

\section{Introduction}
Superconductivity in MgB$_{2}$ 
has attracted much attention since its discovery by Nagamatsu {\em et al.} \cite{nagamatsu01} 
Besides its high transition temperature 
$T_{c}\cong 40K$ for a phonon-mediated pairing mechanism, \cite{phonon} 
it has a striking novel feature that approximately 
two different energy gaps open on different pieces of 
the Fermi surface.
Accompanied with theoretical predictions, \cite{twogap1,twogap2,twogap3,twogap4} 
this multi-gap structure has been established within a short period by specific heat 
experiments, \cite{sph1,sph2,sph3,sph4,sph5,sph6,sph7}
point-contact spectroscopy, \cite{poc1,poc2,poc3} and scanning tunneling microscope. 
\cite{stm1,stm2,stm3}
The larger gap opens on cylindrical Fermi surfaces of the $\sigma$-band
with $2\Delta_{\sigma}(0)/k_{B}T_{c}=3.6-4.5$, whereas 
the smaller one has the size 
$2\Delta_{\pi}(0)/k_{B}T_{c}=1.1-1.5$
on three dimensional 
Fermi surfaces of the $\pi$-band. 

Impurity effects of this material are very interesting to study.
According to Anderson's theorem for classic $s$-wave superconductors, \cite{anderson,abrikosov} 
nonmagnetic impurities do not affect superconducting properties 
in zero magnetic field. 
However, it was shown later by Markowitz and Kadanoff\cite{MK63}
that $T_{c}$ is actually reduced in the presence of gap anisotropy
and impurity scattering.
An application to a two-band model is due to
Golubov and Mazin. \cite{golubov}
Indeed, they predicted
a rather drastic decrease of $T_{c}$ due to \linebreak the interband impurity scattering.
 They also found that, as the interband scattering increases, 
 the density of states changes from the two-gap structure inherent to the two-band 
model to the conventional single-gap structure.
This reduction of $T_{c}$ has been confirmed recently by a couple of experiments.
Wang {\em et\:al.}\cite{sph7}
 measured the specific heat of polycrystalline MgB$_{2}$ after irradiation.
They found both suppression of $T_{c}$ and a tendency towards a \linebreak single-gap structure
 as the scattering is increased by irradiation.
Lee {\em et\:al.}\cite{imp1}
 clarified  the possibility of complete suppression of superconductivity 
by replacing B in MgB$_{2}$ by C.
However, few quantitative calculations have been performed on the impurity effects 
based on a realistic model for MgB$_{2}$.\cite{Mazin02}
For example, the specific heat has been calculated by Choi {\em et\:al.}\cite{twogap2}
and Golubov {\em et\:al.}\cite{sph6} based on 
the clean-limit Eliashberg equation to obtain excellent quantitative agreements.
 However, no detail study has been performed for the specific heat 
from clean to dirty limits based on
a microscopic model.

With these observation, we investigate nonmagnetic impurity effects in MgB$_{2}$.
We thereby clarify impurity-concentration dependence of the quasiparticle 
density of states and the specific heat,
choosing the parameters in the model suitable for MgB$_{2}$.
Section II gives the formulation. Section III presents calculated results.
Section IV summarizes the paper.
We put $\hbar\!=\! k_{\rm B}\!=\! 1$ throughout. 

\section{formulation}
\subsection{Quasiclassical equations}
We start from the Eilenberger equations\cite{eil} 
for the Suhl-Matthias-Walker model\cite{suhl} with impurities,
which form one of the most convenient frameworks to study impurity effects of the
two-band model. 
We here adopt the formulation on the real energy axis instead of using Matsubara frequencies,
which has an advantage that the free-energy functional can be defined unambiguously.

The Eilenberger equation on the real energy axis is given for the uniform case by
\begin{equation}
\left(-i\varepsilon + 
\sum_{\beta} \frac{\langle g_{\beta}^{{\rm R}}\rangle}
{2{\tau_{\alpha \beta }}} \right)
f_{\alpha}^{{\rm R}} 
=\left(\Delta_{\alpha} + 
\sum_{\beta} \frac{\langle f_{\beta}^{{\rm R}}  \rangle}
{2{\tau_{\alpha \beta }}}  \right)
g_{\alpha}^{{\rm R}} \: .  
\label{eq:Eil re}
\end{equation}
Here $f_{\alpha}^{{\rm R}}$$\equiv$$f_{\alpha}^{{\rm R}}( \varepsilon, {\bf k}_{{\rm F}} )$ and 
$g_{\alpha}^{{\rm R}}$$\equiv$$g_{\alpha}^{{\rm R}}( \varepsilon, {\bf k}_{{\rm F}} )$ 
are retarded quasiclassical Green's functions specified by the band index
$\alpha$ $(=\!\sigma,\pi)$ and the Fermi wave vector ${\bf k_{{\rm F}}}$.
They are connected by
$g_{\alpha}^{{\rm R}} = (1-f_{\alpha}^{{\rm R}}f_{\alpha}^{{\rm R}\dagger})^{1/2}$
with $f_{\alpha}^{{\rm R}\dagger}(\epsilon, {\bf k_{{\rm F}}})$
=$f_{\alpha}^{{\rm R}\ast}(-\epsilon, -{\bf k_{{\rm F}}})$.
The symbol $\langle\cdots\rangle$ denotes the average over the Fermi surface for the relevant band
with $\langle 1 \rangle=1$.
The quantity $\tau_{\alpha\beta}$  is the relaxation time for the nonmagnetic impurity scattering 
from $\alpha$- to $\beta$-band;
they satisfy 
$\frac{1}{\tau_{\alpha\beta}}=\frac{N_{\beta}(0)}{N_{\alpha}(0)}\frac{1}{\tau_{\beta\alpha}}$, 
where $N_{\alpha}(0)$ is the normal-state density of states at the Fermi energy
for the $\alpha$-band. 
Finally, $\Delta_{\alpha}$ is determined self-consistently
by
\begin{equation}
\Delta_{\alpha} = \sum_{\beta}\frac{\lambda_{\alpha \beta}}{2i}
\int_{-\varepsilon_{c}}^{\varepsilon_{c}}d\varepsilon \phi(\varepsilon)
\langle f_{\beta}^{{\rm R}}( \varepsilon, {\bf k}_{{\rm F}} ) \rangle \:,
\label{eq:gap re}
\end{equation}
where 
$\varepsilon_{c}$ is the cutoff energy, $\phi(\varepsilon)\!\equiv\!\tanh(\varepsilon/2T)$,
and $\lambda_{\alpha \beta}$ is dimensionless coupling constant with
$\lambda_{\alpha\beta} = \frac{N_{\beta}(0)}{N_{\alpha}(0)} \lambda_{\beta\alpha}$. 

The functional for the free-energy difference corresponding to 
Eqs.  (\ref{eq:Eil re}) and (\ref{eq:gap re}) are given by
\begin{eqnarray}
&&\hspace{-3mm} F_{s}-F_{n} 
\nonumber \\
=&&\hspace{-3mm} 
 \sum_{\alpha(\neq\beta)} N_{\alpha}(0)\biggl\{ 
 \frac{|\Delta_{\alpha}|^{2}}{\lambda_{\alpha\alpha}} 
-\frac{1}{2i} \int_{-\varepsilon_{c}}^{\varepsilon_{c}}d\varepsilon \phi(\varepsilon) 
 \langle I_{\alpha}(\varepsilon ) \rangle 
\nonumber \\
&&\hspace{-3mm}
-\frac{\lambda_{\alpha\beta}}{\lambda_{\alpha\alpha}}
 \frac{1}{2i} \int_{-\varepsilon_{c}}^{\varepsilon_{c}}d\varepsilon \phi(\varepsilon) 
 \biggl[ \Delta_{\alpha}^{\ast}\langle f_{\beta}^{{\rm R}}(\varepsilon )\rangle 
+\Delta_{\alpha}\langle f^{{\rm R}\dagger}_{\beta}(\varepsilon ) \rangle \biggr]  
\nonumber \\
&& \hspace{-3mm}
+\frac{1}{(2i)^{2}}\int_{-\varepsilon_{c}}^{\varepsilon_{c}} 
 \int_{-\varepsilon_{c}}^{\varepsilon_{c}}d\varepsilon d\varepsilon^{\prime}
 \phi(\varepsilon)  
 \phi(\varepsilon')
\nonumber \\
&& \hspace{5mm} 
 \times \biggl[ \frac{\lambda_{\alpha\beta}^{2}}{\lambda_{\alpha\alpha}}
                \langle f_{\beta}^{{\rm R}\dagger}(\varepsilon ) \rangle
                \langle f_{\beta}^{{\rm R}}(\varepsilon^{\prime} ) \rangle 
               +\lambda_{\alpha\beta}
                \langle f_{\beta}^{{\rm R}\dagger}(\varepsilon )\rangle
                \langle f_{\alpha}^{{\rm R}}(\varepsilon^{\prime} )\rangle 
        \biggr] 
\nonumber \\
&& \hspace{-3mm} 
-\frac{1}{2i}\int_{-\varepsilon_{c}}^{\varepsilon_{c}}\!\!d\varepsilon \phi(\varepsilon) 
\nonumber \\
&& \hspace{5mm} 
\times \biggl[ \frac{\langle f_{\alpha}^{{\rm R}\dagger}(\varepsilon) \rangle 
      \langle f_{\beta}^{{\rm R}}(\varepsilon)\rangle
 \!+\!\langle g_{\alpha}^{{\rm R}}(\varepsilon) \rangle 
      \langle g_{\beta}^{{\rm R}}(\varepsilon) \rangle  
 \!-\!1}{2\tau_{\alpha\beta}} \biggr]
\biggr\} \:, \label{eq:free}
\nonumber \\
\end{eqnarray}
with
\begin{eqnarray}
I_{\alpha} &=&
\Delta_{\alpha}^{\ast} f_{\alpha}^{{\rm R}} + \Delta_{\alpha} f_{\alpha}^{{\rm R}\dagger} 
-2i\varepsilon ( g_{\alpha}^{{\rm R}}-1  ) 
\nonumber \\
&&+ \frac{f_{\alpha}^{{\rm R}}\langle f_{\alpha}^{{\rm R}\dagger} \rangle 
  + \langle f_{\alpha}^{{\rm R}}\rangle f_{\alpha}^{{\rm R}\dagger}}{4\tau_{\alpha\alpha}}
  + \frac{g_{\alpha}^{{\rm R}}\langle g_{\alpha}^{{\rm R}} \rangle-1}{2\tau_{\alpha\alpha}}\:. 
\end{eqnarray}
Equation (\ref{eq:free}) is a direct extension of Eilenberger's free-energy functional
\cite{eil} to the two-band model.
Indeed, variations of $F_{s}-F_{n}$ 
with respect to $f^{{\rm R}\dagger}_{\alpha}$ and 
$\Delta^{\ast}_{\alpha}$ 
lead to Eqs.\ (\ref{eq:Eil re}) and (\ref{eq:gap re}), respectively.

The entropy is obtained from this free-energy functional
by $S_{s}=S_{n}-\partial (F_{s}-F_{n})/\partial T$.
Noting the stationarity of $F_{s}-F_{n}$ with respect to $f_{\alpha}^{{\rm R}\dagger}$
and $\Delta_{\alpha}^{\ast}$, we only have to differentiate with respect to the explicit
temperature dependence in $\phi(\epsilon)=\tanh(\epsilon/2T)$. We thereby obtain an explicit
analytic expression for the entropy as
\begin{eqnarray}
&& \hspace{-5mm}S_{s}=S_{n} 
+\sum_{\alpha(\neq\beta)} N_{\alpha}(0)
\frac{1}{2i}\int_{-\varepsilon_{c}}^{\varepsilon_{c}}
d\varepsilon \frac{\partial \phi(\varepsilon)}{\partial T} 
\biggl\{ \langle I_{\alpha}(\varepsilon) \rangle
\nonumber \\
&& \hspace{-3mm} 
+\frac{\langle f_{\alpha}^{{\rm R}\dagger}(\varepsilon ) \rangle 
      \langle f_{\beta}^{{\rm R}}(\varepsilon )\rangle
\!+\! \langle g_{\alpha}^{{\rm R}}(\varepsilon ) \rangle
      \langle g_{\beta}^{{\rm R}}(\varepsilon ) \rangle  
\!-\!1}{2\tau_{\alpha\beta}} 
\biggr\} \:. \label{eq:ent}
\end{eqnarray}
In deriving this expression, we have used Eq.\ (\ref{eq:gap re}).
Finally, the specific heat is calculated by numerically differentiating Eq.\ (\ref{eq:ent}) as 
\begin{equation}
C_{s}=C_{n}+T\frac{d(S_{s}-S_{n})}{dT}\:.  \label{eq:spe}
\end{equation}
Equations  (\ref{eq:ent}) and (\ref{eq:spe}) form a convenient and efficient
starting point to calculate the specific heat for various impurity concentrations.

There is a disadvantage in the coupled self-consistency equations (\ref{eq:Eil re})
and (\ref{eq:gap re}) that they may not be very stable numerically. 
However, it can be removed when $\varepsilon_{c}$ is much larger than both $\Delta_{\alpha}$
and $1/\tau_{\alpha\beta}$. This condition is well satisfied in MgB$_{2}$
where $\varepsilon_{c}$ corresponds to the Debye energy $ \omega_{\rm D}\!\sim\!1000K$.
\cite{sph2}
Then using the asymptotic property 
$f_{\alpha}^{{\rm R}}\!\rightarrow\! \Delta_{\alpha}/(-i\varepsilon)$ 
as $|\varepsilon|\!\rightarrow\!\infty$, Eq.\ (\ref{eq:gap re}) is transformed as
\begin{eqnarray}
\Delta_{\alpha} &\!=\!& \sum_{\beta}\frac{\lambda_{\alpha \beta}}{2i} 
\int_{-\varepsilon_{c}}^{\varepsilon_{c}}\!d\varepsilon \phi(\varepsilon)
\biggl[\langle f_{\beta}^{{\rm R}}( \varepsilon, {\bf k}_{{\rm F}} ) \rangle 
\!-\! \frac{\Delta_{\beta}}{-i\varepsilon} \biggr]
\nonumber \\
&& + \sum_{\beta}\frac{\lambda_{\alpha \beta}}{2} 
\int_{-\varepsilon_{c}}^{\varepsilon_{c}}\!d\varepsilon \phi(\varepsilon)
\frac{\Delta_{\beta} }{ \varepsilon }
\nonumber \\
&=& 2\pi T \sum_{\beta}\lambda_{\alpha \beta}\sum_{n=0}^{\infty} 
\biggl[\langle f_{\beta}( \varepsilon_{n}, {\bf k}_{{\rm F}} ) \rangle 
\!-\! \frac{\Delta_{\beta}}{\varepsilon_{n}} \biggr] \nonumber \\
&&+ \sum_{\beta}\lambda_{\alpha \beta}\Delta_{\beta} \ln\left(\frac{2e^{\gamma}}{\pi}
\frac{\varepsilon_{c}}{T} \right) \:,
\label{eq:gap im}
\end{eqnarray}
where $\gamma=0.577$ is the Euler constant, $\varepsilon_{n}\!\equiv\!(2n\!+\! 1)\pi T$ 
is the Matsubara frequency, and $f_{\alpha}(\varepsilon_{n})
\!=\! f_{\alpha}^{{\rm R}}(i\varepsilon_{n})$.
The limit $\varepsilon_{c}\!\rightarrow\!\infty$ has been taken safely 
in the first integral
to transform the integration into the summation over Matsubara frequencies.
Equation (\ref{eq:gap im}) tells us that $\Delta_{\alpha}$ has no angular dependence
within the band.
It then follows that $f_{\alpha}$ neither has any angular dependence, so that 
$\langle f_{\alpha}\rangle \!=\! f_{\alpha}$ and $\langle g_{\alpha}\rangle \!=\! g_{\alpha}$.
Hence Eq.\ (\ref{eq:Eil re}) with $\varepsilon\!\rightarrow\!i\varepsilon_{n}$
is simplified into
\begin{equation}
\left(\varepsilon_{n} + \frac{g_{\beta} }{2{\tau_{\alpha \beta }}}  \right)
f_{\alpha} = 
\left(\Delta_{\alpha} + \frac{f_{\beta} }{2{\tau_{\alpha \beta}}}  \right)
g_{\alpha}\hspace{5mm} (\beta\neq\alpha) \:.  \label{eq:Eil im imp}
\end{equation}
Thus, the intraband scattering does not affect superconductivity in zero magnetic field at all,
in agreement with Anderson's theorem.\cite{anderson}
Equation (\ref{eq:Eil im imp}) could be presented at the beginning in place 
of Eq.\ (\ref{eq:Eil re}). However, Eqs.\ (\ref{eq:Eil re})-(\ref{eq:free}) 
have an advantage that they could easily be extended to nonuniform systems by simply adding
gradient terms.\cite{eil,Kita03}

Equations (\ref{eq:gap im}) and (\ref{eq:Eil im imp}) enable us to obtain $\Delta_{\alpha}$
from calculations on Matsubara frequencies, which are numerically more stable than
Eqs.\ (\ref{eq:Eil re}) and (\ref{eq:gap re}). Once $\Delta_{\alpha}$ is fixed in this way,
the density of states is calculated by solving Eq. (\ref{eq:Eil re}) as
\begin{equation}
N(\epsilon )=
\sum_{\alpha}N_{\alpha}(0){\rm Re}g_{\alpha}^{{\rm R}}(\epsilon)\:. \label{eq:dos}
\end{equation}

\subsection{Transition temperature}

We now derive the $T_{c}$ equation valid at all impurity concentrations.
To this end, we expand $f_{\alpha}$ and  $g_{\alpha}$ up to first order in $\Delta_{\beta}$ 
as $f_{\alpha}\!=\!f_{\alpha}^{(1)}$ and $g_{\alpha}\!=\! 1$.
Substituting them into Eq.\ (\ref{eq:Eil im imp}), we obtain $f_{\alpha}^{(1)}$ as
\begin{equation}
f_{\alpha}^{(1)}=\sum_{\beta}h_{\alpha\beta}\Delta_{\beta} \, ,
\label{f^(1)}
\end{equation}
where $h_{\alpha\beta}$ is defined by
\begin{equation}
h_{\alpha\beta} = \delta_{\alpha\beta}\left(
\frac{n_{\alpha}}{\varepsilon_{n}} 
+\frac{ 1\!-\! n_{\alpha} }{ \varepsilon_{n} \!+\! \frac{1}{\tau} } \right)+
(1\!-\!\delta_{\alpha\beta})
\left(\frac{n_{\beta} }{\varepsilon_{n}} 
- \frac{n_{\beta} }{ \varepsilon_{n} \!+\! \frac{1}{\tau} } \right) ,
\end{equation}
with
\begin{equation}
n_{\alpha} \equiv \frac{N_{\alpha}(0)}{N_{\alpha}(0)+N_{\beta}(0)}\:
,\hspace{5mm}  \frac{1}{\tau}\equiv \frac{1}{2\tau_{\alpha\beta}}+\frac{1}{2\tau_{\beta\alpha}}\:. 
\end{equation}
Substituting Eq.\ (\ref{f^(1)}) into 
Eq. (\ref{eq:gap im}), we obtain the condition for a nontrivial solution as
\begin{equation}
\det[{\bf 1}-H]=0\:,\label{eq:det}
\end{equation}
where the matrix $H$ is defined by
\begin{eqnarray}
H_{\alpha\beta}&\!=\!&
\sum_{\gamma}\lambda_{\alpha\gamma}\biggl\{2\pi T_{c}\sum_{n=0}^{\infty} 
\biggl[h_{\gamma\beta}(\varepsilon_{n})
\!-\!\frac{\delta_{\gamma\beta}}{\varepsilon_{n}} \biggr] \nonumber \\
&&+\delta_{\gamma\beta}\ln\left(\frac{2e^{\gamma}}{\pi}\frac{\varepsilon_{c}}{T_{c}} \right)
\biggr\} \:.\label{eq:ele}
\end{eqnarray}
By solving Eq.\ (\ref{eq:det}), $T_{c}$ is obtained for an arbitrary $\tau$. 
Notice that all the summations in Eq.\ (\ref{eq:ele}) can be expressed in terms of 
the digamma function $\psi(x)$.

When $T_{c}\tau\!\ll\! 1$, Eq.\ (\ref{eq:det}) can be solved explicitly by using 
the asymptotic expression $\psi(x)\!\sim\! \ln x$ ($x\! \rightarrow\! \infty$) as
\begin{equation}
T_c = \frac{2e^{\gamma}}{\pi}\varepsilon_{c}\exp\!\!\left[
\frac{\displaystyle \sum_{\alpha(\neq\beta)}
(\lambda_{\alpha\beta}n_{\alpha}\!-\!\lambda_{\alpha\alpha}n_{\beta}) 
\ln({\varepsilon_{c} \tau} )\!+\!1}
{\displaystyle
(\lambda_{\sigma\sigma}\lambda_{\pi\pi}\!-\!\lambda_{\sigma\pi}\lambda_{\pi\sigma})
\ln({\varepsilon_{c} \tau} )
\!-\!\sum_{\alpha}\lambda_{\alpha}n_{\alpha} }\right] .
\label{eq:Tc apx}
\end{equation}
with $\lambda_{\alpha}\!\equiv\!\sum_{\beta}\lambda_{\alpha\beta}$.
This expression is useful to see whether $T_{c}$ is suppressed completely or not
as the interband scattering increases.

\subsection{Density of states in the dirty limit}

We now derive an analytic expression for the density of states
in the dirty limit of $T_{c}\tau\!\ll\! 1$. 
In this case with $\varepsilon\!\alt\!T_{c}$, 
we can neglect the first terms on both sides of Eq.\ (\ref{eq:Eil re}).
It then follows that $f_{\alpha}^{{\rm R}}/g_{\alpha}^{{\rm R}}
\!=\! f_{\beta}^{{\rm R}}/g_{\beta}^{{\rm R}}$, or equivalently, 
$f^{{\rm R}}_{\alpha}=f^{{\rm R}}_{\beta} \equiv f^{\rm R}$ and
$g_{\alpha}^{{\rm R}}=g_{\beta}^{{\rm R}}\equiv g^{\rm R}$.
The quantities $f^{\rm R}$ and $g^{\rm R}$ are obtained easily as
\begin{equation}
f^{\rm R}(\varepsilon )\!=\!\frac{\bar{\Delta}}{\sqrt{\bar{\Delta}^{2}-\varepsilon_{+}^{2}}}\:,
\hspace{5mm}
g^{\rm R}(\varepsilon )\!=\!\frac{-i\varepsilon}{\sqrt{\bar{\Delta}^{2}-\varepsilon_{+}^{2}}}\:,
\end{equation}
where $\varepsilon_{+} \!\equiv\!\varepsilon\!+\!i 0_{+}$, and $\bar{\Delta}$ is defined by
\begin{equation}
\bar{\Delta} \equiv n_{\sigma}\Delta_{\sigma} + n_{\pi}\Delta_{\pi}\:.
\label{eq:DeltaBar}
\end{equation}
It hence follows that the density of states $\propto\!{\rm Re}\, g^{\rm R}$ 
has the conventional BCS structure, diverging at
$\varepsilon\!=\! \pm \!\bar{\Delta}$. 
The corresponding pair potential is determined by
\begin{equation}
\Delta_{\alpha} = \frac{\lambda_{\alpha}}{2i}\int_{-\varepsilon_{c}}
^{\varepsilon_{c}} d\varepsilon \phi(\varepsilon) f^{\rm R}(\varepsilon)\:,
\end{equation}
as already pointed out by Golubov and Mazin.\cite{golubov}
Thus, the ratio of the two pair potentials is given by
$\Delta_{\sigma}/\Delta_{\pi}\!=\!\lambda_{\sigma} /
\lambda_{\pi}$. 

\subsection{Numerical procedures}

There are five parameters in the model:
\begin{equation}
\lambda_{\sigma\sigma},\hspace{3mm} \lambda_{\sigma\pi},\hspace{3mm} 
\lambda_{\pi\pi},\hspace{3mm} \varepsilon_{c},
\hspace{3mm} N_{\sigma}(0)/N_{\pi}(0).
\end{equation}
They are fixed so as to reproduce experiments on MgB$_{2}$
as closely as possible.
To be more specific, the ratio $N_{\sigma}(0)/N_{\pi}(0)$ is set equal to $0.72$
following an electronic-structure calculation.\cite{belashchenko}
As for the coupling constants, we use Eq.\ (\ref{eq:gap re}) in the
clean limit at $T=0$, $T_{c}$ which yields
\begin{subequations}
\begin{equation}
1\!=\!\lambda_{\sigma\sigma} \ln\left(\frac{2\varepsilon_{c}}{\Delta_{\sigma}}\right)
\!+\!\lambda_{\sigma\pi} \frac{\Delta_{\pi}}{\Delta_{\sigma}}
\left[ \ln\left(\frac{2\varepsilon_{c}}{\Delta_{\sigma}}\right)
\!-\!\ln\left(\frac{\Delta_{\pi}}{\Delta_{\sigma}}\right) \right] \, ,
\end{equation}
\begin{equation}
\frac{\Delta_{\pi}}{\Delta_{\sigma}}
\!=\!\lambda_{\pi\sigma} \ln\left(\frac{2\varepsilon_{c}}{\Delta_{\sigma}}\right)
\!+\!\lambda_{\pi\pi} \frac{\Delta_{\pi}}{\Delta_{\sigma}}
\left[ \ln\left(\frac{2\varepsilon_{c}}{\Delta_{\sigma}}\right)
\!-\!\ln\left(\frac{\Delta_{\pi}}{\Delta_{\sigma}}\right) \right] \, ,
\end{equation}
and
\begin{equation}
\ln\left(\frac{2{\rm e}^{\gamma}\varepsilon_{c}}{\pi T_{c}}\right)
\!=\! \frac{ \lambda_{\sigma\sigma}\!+\lambda_{\pi\pi}\!+\!\sqrt{
(\lambda_{\sigma\sigma}\!-\lambda_{\pi\pi})^{2}\!+\!4\lambda_{\sigma\pi}\lambda_{\pi\sigma}}}
{2(\lambda_{\sigma\sigma}\lambda_{\pi\pi}\!-\!\lambda_{\sigma\pi}\lambda_{\pi\sigma})} \, ,
\end{equation}
\end{subequations}
respectively.
These equations are used to eliminate $\lambda_{\alpha\beta}$ in favor of 
$\Delta_{\sigma 0}/\Delta_{\pi 0}$, $T_{c}/\Delta_{\sigma 0}$, and 
$\varepsilon_{c}/\Delta_{\sigma 0}$, where $\Delta_{\sigma 0,\pi 0}$
denotes
$\Delta_{\sigma,\pi}(T\!=\! 0)$
in the clean limit.

\begin{table}[t]
\caption{\label{coupling}Coupling constants}
\begin{ruledtabular}
\begin{tabular}{l|llll}
   & $\lambda_{\sigma\sigma}$ & $\lambda_{\sigma\pi}$&$\lambda_{\pi\sigma}$ &$\lambda_{\pi\pi}$ \\
\hline
 Case A & $2.45\times10^{-1}$   & $5.95\times10^{-2}$   &$4.28\times10^{-2}$  &$1.10\times10^{-1}$\\
\hline
 Case B & $2.25\times10^{-1}$   & $2.35\times10^{-1}$   &$1.69\times10^{-1}$  &$-1.22\times10^{-1}$\\
\end{tabular}
\end{ruledtabular}
\end{table}

\begin{figure}[t]
\begin{center}
\includegraphics[width=0.87\linewidth]{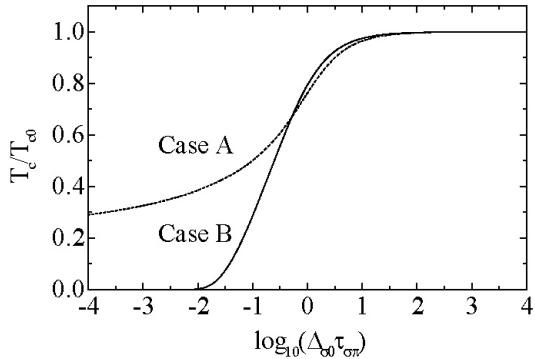}
\caption{Transition temperature as a function of $\log_{10}(\Delta_{\sigma 0}\tau_{\sigma\pi})$}
\label{fig:Tc}
\end{center}
\end{figure}

\begin{figure}[b]
\begin{center}
\includegraphics[width=0.87\linewidth]{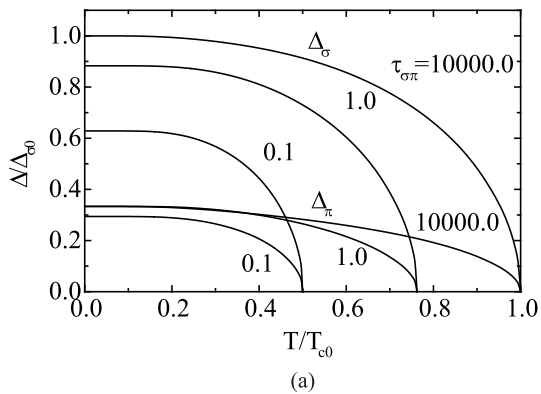}
\includegraphics[width=0.87\linewidth]{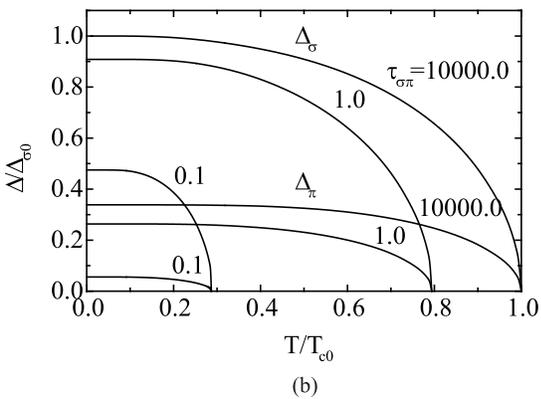}
\caption{The pair potentials as a function of $T/T_{c0}$ for three
different impurity concentrations. (a) Case A; (b) Case B.
Here $\tau_{\sigma\pi}$ is given in units of $\Delta^{-1}_{\sigma 0}$ }
\label{fig:Delta}
\end{center}
\end{figure}

\begin{figure}[t]
\begin{center}
\includegraphics[width=0.87\linewidth]{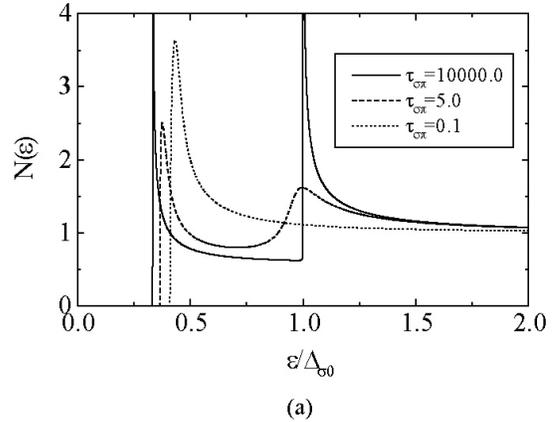}
\includegraphics[width=0.87\linewidth]{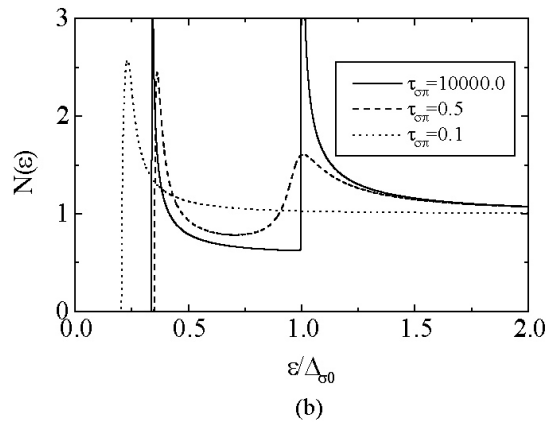}
\caption{Density of states at $T=0.01T_{c}$ for three different $\tau_{\sigma\pi}$.
(a) Case A; (b) Case B.
Here $\tau_{\sigma\pi}$ is given in units of $\Delta^{-1}_{\sigma 0}$ }
\label{fig:DOS}
\end{center}
\end{figure}

We here consider the following two cases:\cite{sph1,sph3,sph5,sph7}
\begin{subequations}
\label{eq:Case12}
\begin{equation}
{\rm Case\:A}: \hspace{1mm}\frac{T_{c}}{\Delta_{\sigma0}}=0.50; 
\hspace{2mm}  \frac{\Delta_{\sigma0}}{\Delta_{\pi0}}=3.00;
\hspace{2mm} \frac{\varepsilon_{c}}{\Delta_{\sigma0}} = 20.0,
\end{equation}
\begin{equation}
{\rm Case\:B}: \hspace{1mm}\frac{T_{c}}{\Delta_{\sigma0}}=0.48; 
\hspace{2mm}  \frac{\Delta_{\sigma0}}{\Delta_{\pi0}}=2.95;
\hspace{2mm} \frac{\varepsilon_{c}}{\Delta_{\sigma0}} = 10.0.
\end{equation}
\end{subequations}
These values are chosen so as to reproduce temperature dependence of the 
observed energy gaps on clean samples\cite{poc1,poc2,poc3,stm2}  as closely as possible.
The corresponding coupling constants are listed in Table I.
It has been found that the whole results are rather insensitive to $\varepsilon_{c}$,
as may be expected.

Numerical calculations have been performed as follows.
First, $f_{\alpha}$ and $g_{\alpha}$ with $f_{\alpha}^{2}\!+\!g_{\alpha}^{2}\!=\! 1$
are expressed conveniently in terms of a single function $a_{\alpha}$ as\cite{Schopohl95,Nagato93}
\begin{equation}
f_{\alpha} =
 \frac{2a_{\alpha}}{1+a_{\alpha}^{2}}, \hspace{10mm}
g_{\alpha} =
 \frac{1-a_{\alpha}^{2}}{1+a_{\alpha}^{2}}
\:.
\label{eq:fg-a}
\end{equation}
Substituting Eq.\ (\ref{eq:fg-a}) and a trial $(\Delta_{\sigma},\Delta_{\pi})$ into it, 
Eq.\ (\ref{eq:Eil im imp}) is transformed into a set 
of nonlinear equations for $a_{\alpha}$,
which may be solved by using one of the standard numerical procedures.\cite{NumRec}
The obtained $f_{\alpha}$ is then substituted into Eq.\ (\ref{eq:gap im})
to find a new $(\Delta_{\sigma},\Delta_{\pi})$. This procedure is repeated until
the convergence is reached.
The pair potentials thereby obtained are then used in Eq.\ (\ref{eq:Eil re}) 
to calculate $f^{{\rm R}}$ and $g^{{\rm R}}$ on the real energy axis.
Finally, those $f^{{\rm R}}$ and $g^{{\rm R}}$ are substituted into 
Eqs.\ (\ref{eq:ent}), (\ref{eq:spe}), and (\ref{eq:dos})
to calculate the specific heat and the density of states.

\section{results}

\begin{figure}[t]
\begin{center}
\includegraphics[width=0.87\linewidth]{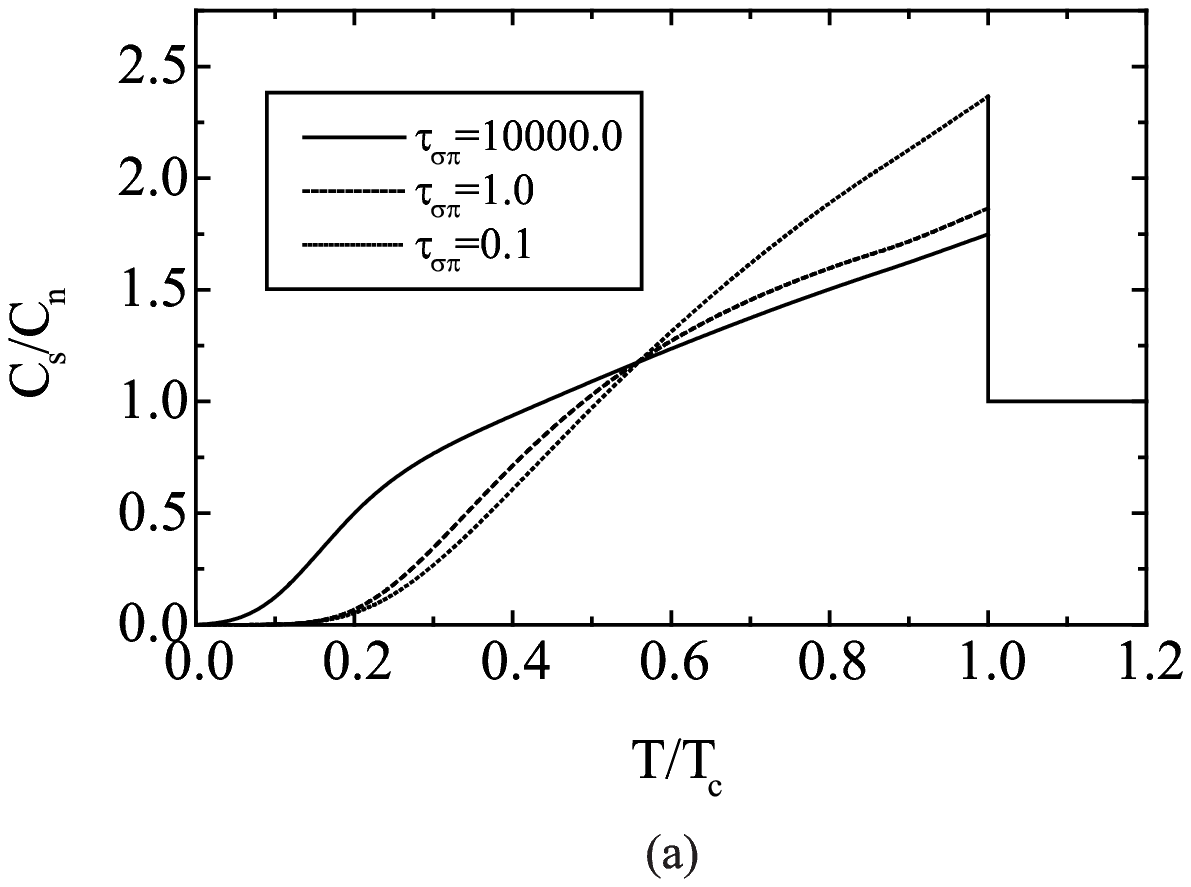}
\includegraphics[width=0.87\linewidth]{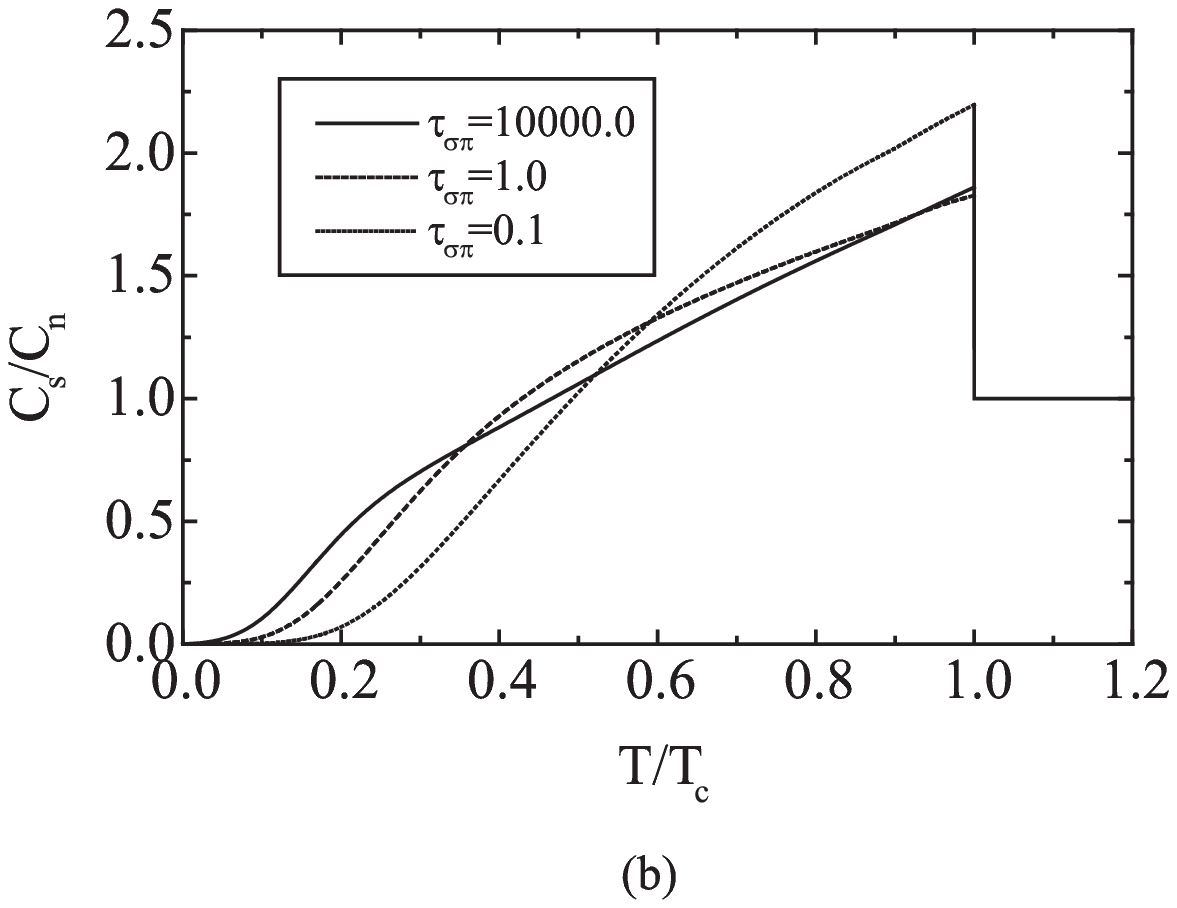}
\caption{Specific heat as a function of $T/T_{c}$ for 
$\tau_{\sigma\pi}\!=\!10000.0$, $1.0$, and $0.1$.
(a) Case A; (b) Case B.
Here $\tau_{\sigma\pi}$ is given in units of $\Delta^{-1}_{\sigma 0}$ }
\label{fig:C}
\end{center}
\end{figure}

Figure \ref{fig:Tc} plots $T_{c}$ as a function of 
the interband scattering specified by $\log_{10}(\Delta_{\sigma 0}\tau_{\sigma\pi})$
for the two cases of Eq.\ (\ref{eq:Case12}). 
Here $T_{c}$ is
normalized by $T_{c0}$ without the interband scattering.
We observe that $T_{c}$ drops steeply around $\tau_{\sigma\pi}\!=\!
\Delta_{\sigma 0}^{-1}$ in both cases.
However,
whereas $T_{c}$ in Case A remains finite for $\tau_{\sigma\pi}\!\rightarrow\! 0$,
$T_{c}$ in Case B decreases to zero at a finite $\tau_{\sigma\pi}$.
This difference can be realized from Eq.\ (\ref{eq:Tc apx}) where the term in the square
bracket passes through negative infinity for $T_{c}\!\rightarrow\!0$.
It is interesting to see experimentally whether $T_{c}$ of MgB$_{2}$ 
is suppressed completely or
not by increasing the interband scattering, although this may not be easy.\cite{Mazin02}

Figure \ref{fig:Delta} shows temperature dependence of the pair potentials 
for (a) Case A and (b) Case B. 
In both cases, the pair potentials 
decrease as $\tau_{\sigma\pi}$ becomes shorter. 
However, the two pair potentials do not approach to a single value
even for $\tau_{\sigma\pi}\!\rightarrow\!0$.
Notice also that the pair potentials are not directly connected with any observable quantities
except in the clean limit.

\begin{figure}[htbp]
\begin{center}
\includegraphics[width=0.9\linewidth]{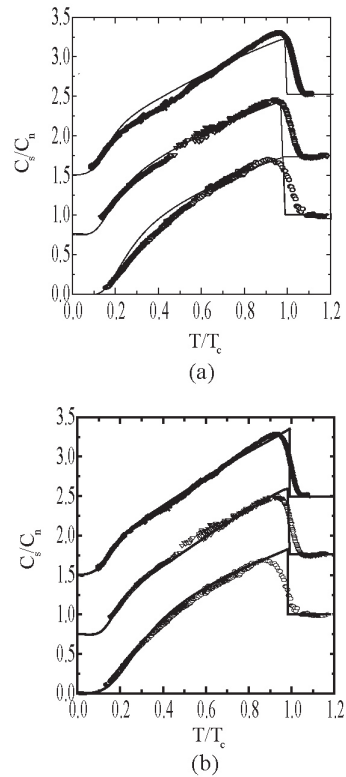}
\caption{Comparisons of theoretical curves 
with specific-heat measurements before irradiation (top), after the first irradiation (middle),
and after the second irradiation (bottom) of Ref.\ \onlinecite{sph7}.
(a) Case A with $\Delta_{\sigma0}\tau_{\sigma\pi}\!=\! 50.0$, $7.0$, $4.0$ 
from top to bottom. (b) Case B $\Delta_{\sigma0}\tau_{\sigma\pi}\!=\! 10000.0$, $30.0$, $1.0$
from top to bottom.}
\label{fig:com}
\end{center}
\end{figure}

Figure \ref{fig:DOS} shows the density of states at $T/T_{c}=0.01$
for (a) Case A and (b) Case B. 
In the clean limit $\Delta_{\sigma0}\tau_{\sigma\pi}\!=\!10000.0$, 
we clearly observe a couple of divergences 
at $\varepsilon\!=\! \Delta_{\sigma}$, $\Delta_{\pi}$.
As the interband scattering 
becomes larger, the divergences are smeared to finite peaks,
which eventually merge into a single peak in the dirty limit
at $\varepsilon\!=\!\bar{\Delta}$ 
given by Eq.\ (\ref{eq:DeltaBar}).
Notice that the excitation threshold is not a monotonic function of 
$\tau_{\sigma\pi}$; for Case A, for example, the threshold is seen to
increase as $\tau_{\sigma\pi}$ becomes smaller.

Figure \ref{fig:C} plots the specific heat as a function of $T/T_{c}$
in (a) Case A and (b) Case B for
$\Delta_{\sigma0}\tau_{\sigma\pi}=10000.0$, $1.0$, $0.1$. 
A shoulder is clearly seen around $T/T_{c}\!=\! 0.2$ for 
$\Delta_{\sigma0}\tau_{\sigma\pi}\!=10000.0$,
corresponding to the divergence at $\varepsilon\!=\!\Delta_{\pi}$ 
in Fig.\ref{fig:DOS}. It gradually disappears as 
$\tau_{\sigma\pi}$ becomes shorter, however,
and we finally have a conventional single-exponential behavior in the dirty limit.

Finally, Fig.\ \ref{fig:com} compares the present theory with
the specific-heat experiment after irradiation 
performed by Wang {\em et al}.\cite{sph7} 
The data points correspond to measurements
before irradiation (top), after the first irradiation (middle), and 
after the second irradiation (bottom), where
the former two are shifted upwards by $1.5C_{n}$ and $0.75C_{n}$, respectively.
The curves of Fig.\ 5(a) are obtained by using the parameters of 
Case A with $\Delta_{\sigma0}\tau_{\sigma\pi}\!=\! 50.0$, $7.0$, $4.0$ 
from top to bottom, respectively.
On the other hand, those of Fig.\ 5(b) correspond to Case B with 
$\Delta_{\sigma0}\tau_{\sigma\pi}\!=\! 10000.0$, $30.0$, $1.0$.
The agreements are good for both cases, especially for Case B.
From these comparisons, we realize that the interband scattering 
after the second irradiation is still not
very strong with $\Delta_{\sigma0}\tau_{\sigma\pi}\!\agt \! 1.0$.
The fact may imply the difficulty of introducing the interband impurity
scattering in MgB$_{2}$.\cite{Mazin02}
This tendency is also seen in the carbon-substituted system Mg(B$_{1-x}$C$_{x}$)$_{2}$
where the two-gap structure is still observed clearly for $x\!\sim\!0.1$.\cite{Matsuda04}
Thus, the large reduction of $T_{c}$ observed in Mg(B$_{1-x}$C$_{x}$)$_{2}$
by Lee {\em et al}.\cite{imp1}
should be attributed not only to the interband scattering alone but also
to a change of the pairing interaction due
to the electronic structure.

\section{Conclusion}
We have studied nonmagnetic impurity effects for MgB$_{2}$
based on the quasiclassical equations of superconductivity
for the Suhl-Matthias-Walker model.
The parameters in the model are fixed so as to reproduce experimental
values for $\Delta_{\pi 0}/\Delta_{\sigma 0}$ and $\Delta_{\sigma 0}/T_{c}$.
The interband impurity scattering tends to reduce the gap anisotropy.
We have clarified how the density of states changes from the two-gap structure
in the clean limit to the single-gap structure in the dirty limit 
with strong interband scattering. 
Especially, there may be cases where the excitation threshold 
increases as the scattering becomes stronger.
Calculated curves for the specific heat agree well with measurements before
and after irradiation.
This comparison has also enabled us to estimate the relaxation time $\tau_{\sigma\pi}$
for the interband scattering. It satisfies $\tau_{\sigma\pi}\!\agt \! 1/\Delta_{\sigma 0}$
even after the second irradiation,
implying the difficulty of introducing the interband scattering in this system.

\acknowledgements
This work is supported by a Grant-in-Aid for Scientific Research from the Ministry
of Education, Culture, Sports, Science, and Technology of Japan.

\end{document}